\documentclass{elsart}
\usepackage{epsfig}
\usepackage{amssymb}

\begin{document}

\begin{frontmatter}

\title{ Isoscaling in statistical fragment emission in an extended compound nucleus model}

\author{W. Ye \corauthref{TH1}},
\author{J. T\~{o}ke} and
\author{W. U. Schr\"{o}der}
\address{Departments of Chemistry and Physics, University of
Rochester, Rochester, New York 14627, USA }

\corauth[TH1]{Present address: Department of Physics,  Southeast University, Nanjing 210096, People's Republic of China }

\begin{abstract}
Based on an extended compound nucleus model, isospin effects in
statistical fragment emission from excited nuclear systems are
investigated.  An experimentally observed scaling behavior of the
ratio of isotope yields $Y_i(N,Z)$ from two similar emitting
sources with different neutron-to-proton ratios is predicted
theoretically, i.e., the relationship of $Y_2/Y_1 \propto
exp(\alpha N + \beta Z)$ is demonstrated.  The symmetry energy
coefficient $C_{sym}$ extracted from the simulation results is
$\sim$ 27 MeV which is consistent with realistic theoretical
estimates and recent experimental data. The influence of the
surface entropy on the isoscaling behavior is discussed in detail.
It is found that although the surface entropy increases the
numercial values of isoscaling parameters $\alpha$ and $\beta$, it
does not affect the isoscaling behavior qualitatively and has only
a minor effect on the extracted symmetry energy coefficient.

\end{abstract}

\begin{keyword}
Isoscaling \sep
Compound nucleus  \sep
Fragment emission

\PACS 25.70.Pq
\sep 24.10.Pa
\sep 21.65.+f
\sep 21.60.Ev

\end{keyword}

\end{frontmatter}

\section{INTRODUCTION}
\vspace{5mm}

The search for a nuclear equation of state (EOS) has been one of the dominant factors driving interest in heavy-ion collisions at intermediate and high energies, ever since such beams have become available. Exploration of the isospin dependence of the EOS is an inseparable part of the task. Originally, the latter studies were confined to use stable projectile beams with only a moderately wide range of isotopes between neutron-poor and neutron-rich nuclei. However, exotic secondary beams of projectile nuclei with extreme neutron-to-proton ratios that have become available recently offer new and
intriguing opportunities to study isospin physics in heavy-ion
collisions \cite{LiSch,Baran,LBA}. These studies have already
resulted in the discovery of several interesting isospin-related
systematics \cite{Sch2,Liu1,Ger,Tsang1}, prompting further efforts
both, in experiments and in theoretical modeling. In particular, a so-called isoscaling analysis of light fragment emission
in the decay of very hot nuclear systems produced in heavy-ion
collisions has attracted attention as a possible tool for deducing the nuclear symmetry energy from the relative fragment yields
\cite{Tsang1,Tsang2,Tsang3,Bot1,Shetty1}.

For strongly damped collisions fragment isotopic yields were found
to follow a ``$Q_{gg}$ systematics'' \cite{Volk,Gelb,Sch3}.
Isotopic scaling, termed ``isoscaling'' \cite{Tsang1,Tsang2,Tsang3},
is observed for other types of reactions such as
evaporation \cite{Tsang1,Tian}, fission \cite{Fri,Ves},
projectile fragmentation \cite{Sou,Ves2}, and multifragmentation
\cite{Tsang1,Liu1,Ger}.  This scaling law refers to a general
exponential relation between the ratios of yields $Y_i(N,Z)$ for
fragments ($N,Z$) emitted from systems which differ only in their
isospin or $(\frac{N_s}{Z_s})_i$. In particular, if two reactions
lead to primary systems $i$ = 1 and 2 having approximately the
same temperature but different isospin, the ratio $R_{21}(N,Z)$ of
the experimental yields of a given fragment $(N,Z)$ emitted from
these systems exhibits an exponential dependence on the fragment
neutron number $N$ and atomic number $Z$ of the form,
\begin{equation}
R_{21}(N,Z)=\frac{Y_{2}(N,Z)}{Y_{1}(N,Z)}=Cexp(\alpha N + \beta
Z).
\end{equation}
Here, $\alpha$ and $\beta$ are the isoscaling parameters and $C$
is an overall normalization constant.

On the theoretical side, isoscaling has been extensively examined
in the antisymmetrical molecular dynamics model \cite{Ono}, a
Boltzmann-Uehling-Uhlenbeck model \cite{Liu1}, a lattice gas model
\cite{Ma1,Das1}, the expanding evaporating source model
\cite{Fri2} and in statistical multifragmentation models
\cite{Tsang2,Tsang3,Bot1,Souza}.  In the present paper it will
be shown how the essential features of isoscaling observed in
experiment are related to the nuclear symmetry energy in an
extended compound nucleus (ECN) model. This model is closely
related to that known from fission studies, and its central notion
is a relatively high entropy (particularly at moderately high
excitation energies) associated with the diffuse nuclear surface
region (as opposed to bulk matter). Application of this ECN model
reveals a new mechanism of nuclear fragmentation caused by the
softening of the diffuse nuclear surface at high excitations
\cite{Toke1}. The ECN model is equivalent to conventional compound
nucleus models at low excitation energy as represented, e.g., by
GEMINI \cite{Charity}, but greatly extends the validity of the
compound nucleus concept towards high excitations \cite{Toke0}.
Therefore, the ECN model provides a unified description of
statistical emission of light particles and fragments in a wide
range of excitation energies.  While the present model is still
somewhat schematic, it is based on a physically transparent
picture and thus provides direct insight into the phenomena of
interest.

\vspace{5mm}
\section{THEORETICAL FRAMEWORK}
\vspace{5mm}

The probability $p$ of emitting a fragment from an equilibrated
compound nucleus (CN) is evaluated using the Weisskopf formalism
\cite{Weis}:
\begin{equation}
p \propto e^{\Delta S} = e^{S_{saddle} - S_{eq}},
\end{equation}
where $S_{eq}$ and $S_{saddle}$ are the entropies for the
equilibrated CN and a saddle-point configuration of touching
spheres, respectively. Within the Fermi gas model, the entropies
for the two configurations can be calculated as
\begin{equation}
S_{eq} = 2 \sqrt{a_{A} E^{*}_{tot}} ,
\end{equation}
and
\begin{equation}
S_{saddle} = S_{res} + S_{frag} = 2 \sqrt{(a_{res}+a_{frag})
E^{*th}_{saddle}} .
\end{equation}
In Eqs. (3) and (4), $a_{A}$, $a_{res}$ and $a_{frag}$ are the
level density parameters of the CN system at equilibrium, of the
residue, and of the fragment, respectively.  The part $S_S$ of the
entropy $S$ of the system, associated with the diffuse surface
domain, has been found \cite{Toke2} to have a pronounced effect on
the fragment emission probability. One has
\begin{equation}
S = S_V + S_S,
\end{equation}
where $S_V$ is the entropy of bulk matter.  Consequently,
the level density parameter $a$ includes volume
and surface terms \cite{Toke4},
\begin{equation}
a = a_V + a_S = \alpha_{V} A + \alpha_{S} A^{2/3} F_{2},
\end{equation}
Here $A$ is the atomic number, $F_2$ is the surface area relative to a
spherical shape.  $E^{*th}_{saddle}$ in
Eq.(4) is the thermal excitation energy of the system in the
saddle-point configuration.  The latter quantity is calculated as
\begin{equation}
E^{*th}_{saddle} = E^{*}_{tot} - V_{saddle},
\end{equation}
where $V_{saddle}$ is the collective saddle-point energy.

For the level density parameter $a$, the parametrization of
T\~{o}ke and Swiatecki \cite{Toke4} was employed with $\alpha_{V}
= 1/14.6$ MeV$^{-1}$ and $\alpha_{S} = 4/14.6$ MeV$^{-1}$.   The
calculations assume saddle-point shapes to be represented by two
touching spheres. $V_{saddle}$ is the difference in deformation
energies for the saddle-point shape and equilibrium-state shape.
It contains contributions from Coulomb, volume, and surface
energies. Volume and surface energies depend on the system isospin
$I$ (defined as $(N-Z)/A$) in a functional form
$-\alpha_{v}(1-\kappa_{v} I^{2}) A$ and $\alpha_{s}(1-\kappa_{s}
I^{2}) A^{2/3}$, respectively.  The parameters are taken from
\cite{Moller} as $\alpha_s = 21.13, \kappa_s = 2.3, \alpha_v =
15.9937, \kappa_v = 1.927 $. The nuclear temperature can be
obtained from the commonly used Fermi-gas model relationship
between the temperature $T$ and the excitation energy of the
system $E^*_{tot}$:
\begin{equation}
T = \sqrt \frac{E^*_{tot}}{a}.
\end{equation}

The present calculations do not account for the effects of an
expansion of the CN prior to its decay, so the nuclear density is
that of the ground state.  Therefore the additional influence of
the nuclear matter density on the extraction of symmetry energy
coefficient is not studied here.  The present work concentrates on
the temperature dependence of the relation between symmetry energy
and isoscaling parameters.

\vspace{5mm}
\section{RESULTS AND DISCUSSION}
\vspace{5mm}

In this work several pairs of equilibrated CN sources with proton
number $Z_s$ = 75 and mass numbers $A_s$ = 165, 175, 185, and 195,
are considered at initial excitation energies of $E^*_{tot}$/A =
2, 3, 4, 5, 6, and 7 MeV/nucleon.  As the ratio $R_{21}(N,Z)$ is
insensitive to sequential decay, it carries information on the
original excited fragments prior to their decay \cite{Sou}. In
particular, it has been found that the values of $\alpha$ and
$\beta$ are not much affected by sequential decay of the primary
fragments \cite{Tsang3}.  These observations justify the neglect
of sequential decay in the following analysis.

The yield ratio $R_{21}(N,Z)$ is constructed using the convention
that index 2 refers to the neutron-rich system and index 1 to the
neutron-poor one.  The observation summarized in Eq. (1) shows
that experimental yield values of $ln(R_{21}(N,Z))$ plotted {\it
vs.} $N$ ($Z$ fixed) or $Z$ ($N$ fixed), produce straight lines.
This feature also emerges from the present MCN model.  As a
demonstration, Fig. 1 shows the theoretical yield ratios
$ln(R_{21}(N,Z))$ plotted {\it vs.} fragment neutron number $N$,
for individually values of $Z$ = $6-9$ (top panel), and {\it vs.}
proton number $Z$, for individually values of $N$ = $6-9$ (bottom
panel).  The ratios have been calculated for the two source pairs
($A_2$ = 175, $A_1$ = 165) and  ($A_2$ = 185, $A_1$ = 165) at an
excitation energy of $E^*_{tot}/A$ = 5 MeV/nucleon. One can see
from Fig.1 that isotope ratios of the same element $Z$, or isotone
ratios of the same $N$, tend to lie on a logarithmic straight
line. The solid lines represent linear fits to each series of $Z$
isotopes and each series of $N$ isotones, respectively. It is
obvious that these fit lines are nearly parallel to each other.
The positive slope in the upper panel of Fig.1 illustrates that
neutron-rich fragments are more easily produced from the more
neutron-rich sources, as can be expected. Analogously, the
negative slope in the bottom panel of Fig.1 illustrates that
proton-rich fragments are more readily produced from the more
proton-rich emitting sources. These are the typical features
associated with ``isoscaling'' behavior showing that the scaling
behavior of the isotope and isotone yield ratios appears naturally
within the present CN model framework.

It was reported \cite{Toke2} that the ratio of a certain fragment
($N,Z$) emission probability from two equilibrated systems can
approximately be written as $P_{2}(N,Z)/P_{1}(N,Z)$ $\propto$
$exp[(V_{1}(N,Z)-V_{2}(N,Z))/T]$, where $V_{i}(i=1,2)$ is of the
same meaning as $V_{saddle}$, namely the interaction energy at the
saddle point. An analysis indicates that in the difference ($V_1 -
V_2$), most terms cancel except for terms that are directly
related to the isospin of the CN system. These isospin-related
terms can be combined as ($\alpha_{v}\kappa_{v} -
\alpha_{s}\kappa_{s}/A^{1/3}$) $(N-Z)^2/A$ = $C_{sym}$ $(N-Z)^2/A$
= $E_{sym}$.  Here $C_{sym}$ (= $\alpha_{v}\kappa_{v} -
\alpha_{s}\kappa_{s}/A^{1/3}$) is symmetry energy coefficient
\cite{Souza2} incorporating both the volume and surface
contributions to the symmetry energy $E_{sym}$ \cite{Tsang3}.
Hence the value of ($V_1 - V_2$) is dominantly determined by the
difference in symmetry energy between the two emitting sources and
their residues. This demonstrates that the origin of the
isoscaling phenomenon found in the present frame can be traced to
the symmetry energy term in the nuclear binding energy, as also
suggested in other work (e.g., \cite{Tsang1}).

The isoscaling parameters $\alpha$ and $\beta$ can be extracted
from fits of model predictions to the data points shown in Fig.1.
An average value of $\alpha$ is calculated over the range 5 $\leq
Z \leq$ 9, an average $\beta$ is calculated over the range 5 $\leq
N \leq$ 9. Figure 2 depicts the dependence of the average
coefficients $\alpha$ and $\beta$ {\it vs.} excitation energy
(left panel) and {\it vs.} inverse of the temperature (right
panel).  One notices that the absolute values of $\alpha$ and
$\mid \beta \mid$ decrease monotonically with excitation energy,
implying that isospin effects decrease with increasing excitation
energy. There is a significant sensitivity to the excitation
energy at low energy, but both the sensitivity to excitation
energy and the overall isospin effect are weakened at very high
excitation.  In addition, it is rather evident that both $\alpha$
and $\beta$ show a linear dependence on $1/T$.

The surface contribution $a_S$ to the level density parameter $a$
appreciably enhances the fragment emission probability via
significant surface entropy effects \cite{Toke2}.  It is therefore
interesting to examine the effects of surface entropy on the
isoscaling phenomenon in the present model. Results are displayed
in Figs.3 and 4, to be compared to Figs.1 and 2 which have been
calculated without surface contribution to the entropy. Evidently,
familiar isoscaling behavior is predicted well in either case. The
linear relationship between isoscaling parameters and inverse
temperature is not affected by the surface contribution either, as
seen from the right panel of Fig.4. However, an account of the
surface term predicts larger values of $\alpha$ and $\beta$
(comparing Fig.4 and Fig.2), implying that the surface entropy
effect on fragment emission becomes stronger with increasing
isospin of the emitting sources, also as expected.

Since the scaling behavior of ratios of fragment isotopic yields
measured in separate nuclear reactions has been utilized to probe
the symmetry energy \cite{Tsang1,Shetty1,Shetty2,Souza2}, in
the following a symmetry energy coefficient is extracted from the
theoretical isoscaling systematically analytical expressions for
$\alpha$ and $\beta$. It is worth noting that, for a given
$E^*_{tot}/A$ these four sources have the same temperature when
the surface contribution to the level density parameter is
neglected.

It has been shown \cite{Tsang3,Bot1} that the isoscaling parameters $\alpha$ and $\beta$
are related to
the symmetry energy coefficient $C_{sym}$ as
\begin{equation}
\alpha = 4\frac{C_{sym}}{T}[(\frac{Z_s}{A_s})^{2}_{1} -
(\frac{Z_s}{A_s})^{2}_{2}]
\end{equation}
and
\begin{equation}
\beta = 4\frac{C_{sym}}{T}[(\frac{N_s}{A_s})^{2}_{1} -
(\frac{N_s}{A_s})^{2}_{2}].
\end{equation}
Equations (9) and (10) have been used to constrain the symmetry
energy coefficient $C_{sym}$ based on experimental data (e.g.
Ref.\cite{Shetty1}).  Figure 5 depicts the product $\alpha \cdot
T$ as a function of $(Z_s/A_s)^{2}_{1} - (Z_s/A_s)^{2}_{2}$ and
$\beta \cdot T$ as a function of $(N_s/A_s)^{2}_{1} -
(N_s/A_s)^{2}_{2}$ for the initial four source pairs at various
excitation energies. All these systems with different source sizes
and isospin asymmetries lie along one single straight line, which
illustrates that the isoscaling parameters $\alpha$ and $\beta$
are not sensitive to the system size.  By fitting the theoretical
data in Fig. 5 with Eqs. (9) and (10), a symmetry energy
coefficient $C_{sym}$ can be obtained to be 27.3 $\pm$ 0.1 MeV,
which is close to the standard liquid drop model value $C_{sym}$ =
25 MeV and also in reasonable agreement with that obtained in
Ref.\cite{Sou}. In that experimental work \cite{Sou}, a value of
$C_{sym}$ = (27.2 $\pm$ 2.2) MeV was deduced based on an
experimental analysis of 25-MeV/nucleon $^{86}$Kr + $^{124,112}$Sn
reactions.  The present analysis demonstrates consistency with
other realistic theoretical estimates \cite{Stei,LBA} and
indicates the validity of an interpretation of isoscaling data in
terms of the symmetry energy.

Although the inclusion of the surface term $a_S$ in the level
density parameter leads to a difference in the temperatures for
the four systems, calculations indicate that for each excitation
energy $E^*_{tot}/A$, the temperature differences among the four
sources are less than 0.1 MeV, so the temperature value of four
systems can be regarded as approximately constant when the
excitation energy per nucleon is fixed. Figure 6 displays the
effect of the surface entropy on the extracted symmetry energy
coefficient.  The value deduced from the theoretical data in Fig.
6 is $C_{sym}$ = (27.6 $\pm$ 0.1) MeV, which is comparable with
that predicted without considering the surface entropy effects.
The reason is that although the surface entropy increases the
values of $\alpha$ and $\beta$, it also decreases the value of
temperature due to an increased level density parameter $a$ [see
Eq. (6)]. Consequently, $\alpha \cdot T$ and $\beta \cdot T$ are
little affected by the surface entropy.  This result indicates
that surface entropy probably has a minor influence on the
extraction of the symmetry energy coefficient. Moreover, it also
implies that the isoscaling observable is a robust probe of the
symmetry energy.

\vspace{5mm}
\section{SUMMARY AND CONCLUSIONS}
\vspace{5mm}

In conclusion, in the framework of an extended compound nucleus
model, isoscaling behavior and its relation to the nuclear
symmetry energy is revealed.  The symmetry energy coefficient is
found to be $C_{sym} \sim$ 27 MeV from an analysis of theoretical
``data'', suggesting that isoscaling data can be interpreted in
terms of the symmetry energy.  In addition, in this work surface
entropy influences on the isoscaling phenomenon have been studied.
The present ECN model approach leaves sufficient room for further
improved treatment of interesting physical effects \cite{Toke2},
such as nuclear expansion.  Such a more detailed analysis is
required to study the evolution of the symmetry energy with the
excitation energy, as has been explored recently also by Shetty
{\it et al.} \cite{Shetty3}.  The model will be utilized to deduce
the effects of nuclear expansion on the isoscaling parameters and
will correspondingly illustrate how to deduce the density
dependence of the symmetry energy in such more general and
realistic scenarios. Work along this direction is in progress.

\vspace{12mm}
{\bf ACKNOWLEDGMENTS}
\vspace{5mm}

 This work is supported by the U.S. Department of Energy Grant
 No. DE-FG02-88ER40414.  The work of W.Y is also partially
 supported by the NSFC under Grant No. 10405007 and China scholarship
 council.  W.Y is also grateful to Rochester University for support and
 hospitality extended to him.

\newpage
\begin{figure}[htb]
\centering
\includegraphics[scale=1.0]{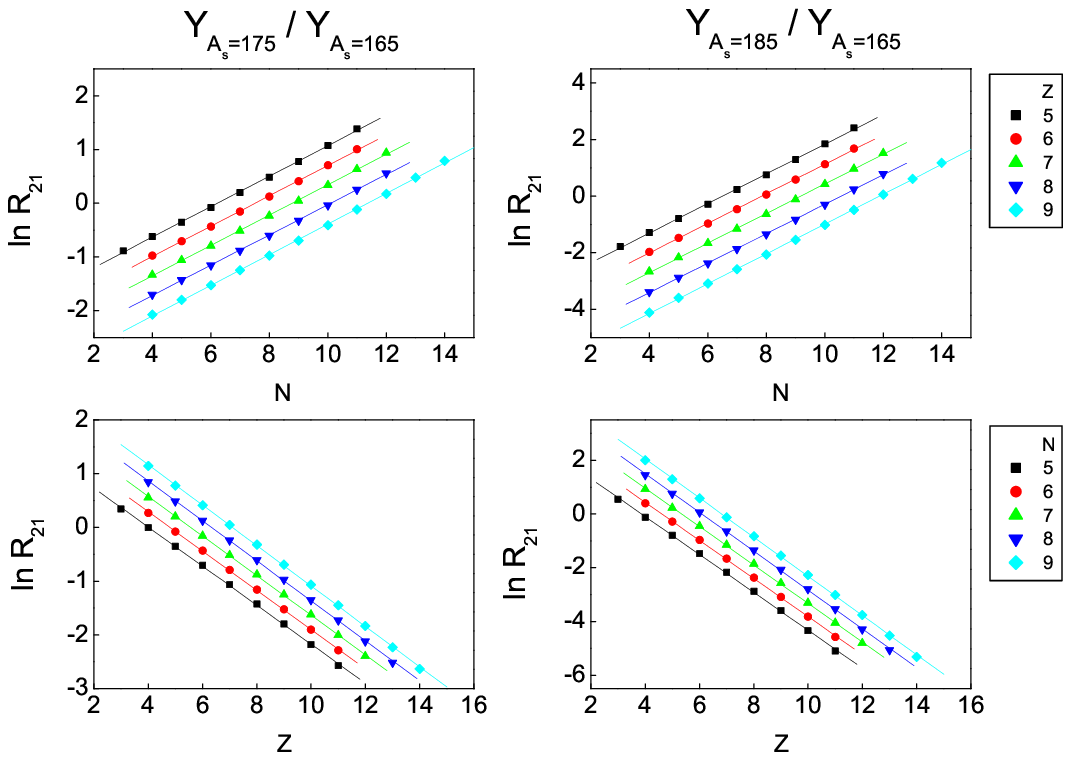}
\caption {\large (Color online) The logarithm of the ratio of
elements $Z$ = 5-9 isotopes yields (top panel) and of $N$ = 6-9
isotones yields (bottom panel) from pairing source $A_s$ = 175 and
$A_s$ = 165 (left column) as well as from pairing source $A_s$ =
185 and $A_s$ = 165 (right column) at excitation energy
$E^*_{tot}/A$ = 5 MeV/nucleon. Here the calculations do not
consider the surface entropy, i.e., not including $a_S$ in the
expression for the level density parameter $a$ [see Eq. (6)].
Solid lines are the linear fitting to the data points. }
\end{figure}

\newpage
\begin{figure}[htb]
\centering
\includegraphics[scale=1.0]{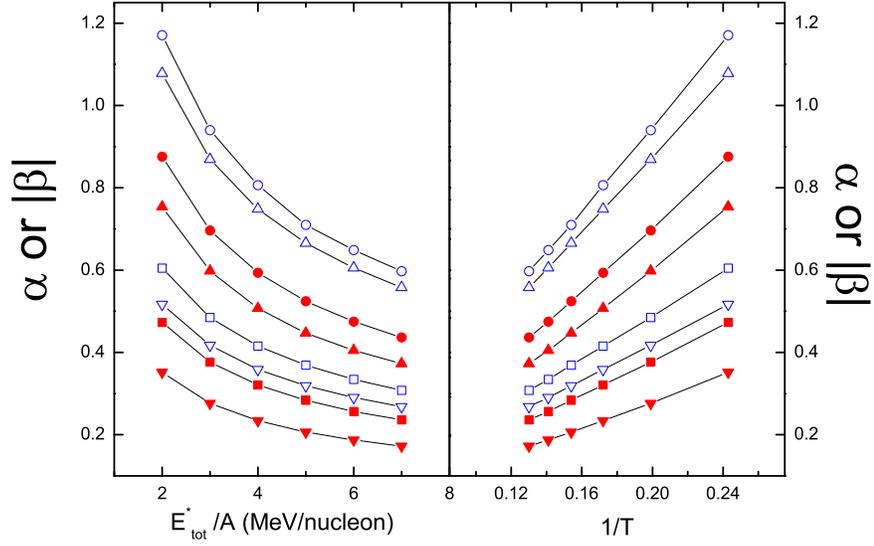}
\caption {\large (Color online) Dependence of isoscaling
parameters $\alpha$ and $\mid \beta \mid$ on excitation energy
(left panel) and the inverse temperature (right panel) for various
source pairs. Symbols in the figures are $\alpha$ (solid symbols)
or $\mid \beta \mid$ (open symbols) from four source pairs
$Y_{A_{s}=175}$/$Y_{A_{s}=165}$ (squares),
$Y_{A_{s}=185}$/$Y_{A_{s}=165}$ (circles),
$Y_{A_{s}=195}$/$Y_{A_{s}=175}$ (up-triangles), and
$Y_{A_{s}=195}$/$Y_{A_{s}=185}$ (down-triangles). Here the
calculations do not consider the surface entropy, i.e., not
including $a_S$ in the expression for the level density parameter
$a$ [see Eq. (6)]. }
\end{figure}

\newpage
\begin{figure}[htb]
\centering
\includegraphics[scale=1.0]{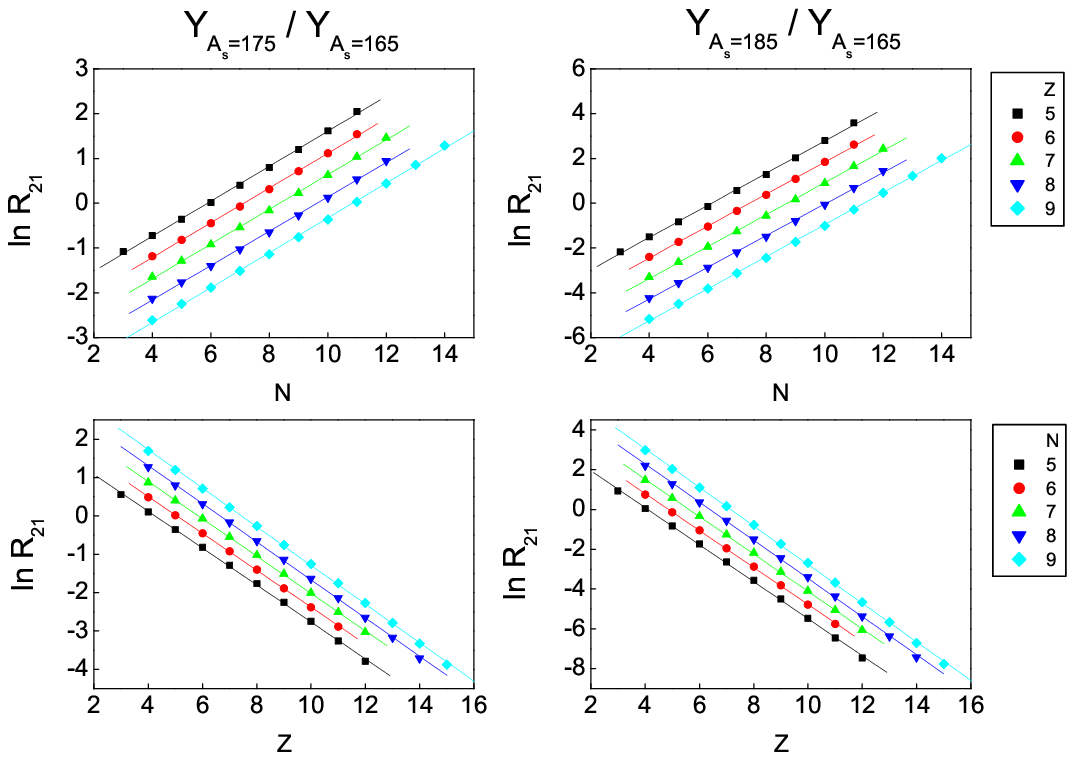}
\caption {\large (Color online)  Same as Fig.1 but consider the
surface entropy effect via inclusion of $a_S$ in the expression
for the level density parameter $a$ [see Eq. (6)]. }
\end{figure}

\newpage
\begin{figure}[htb]
\centering
\includegraphics[scale=1.0]{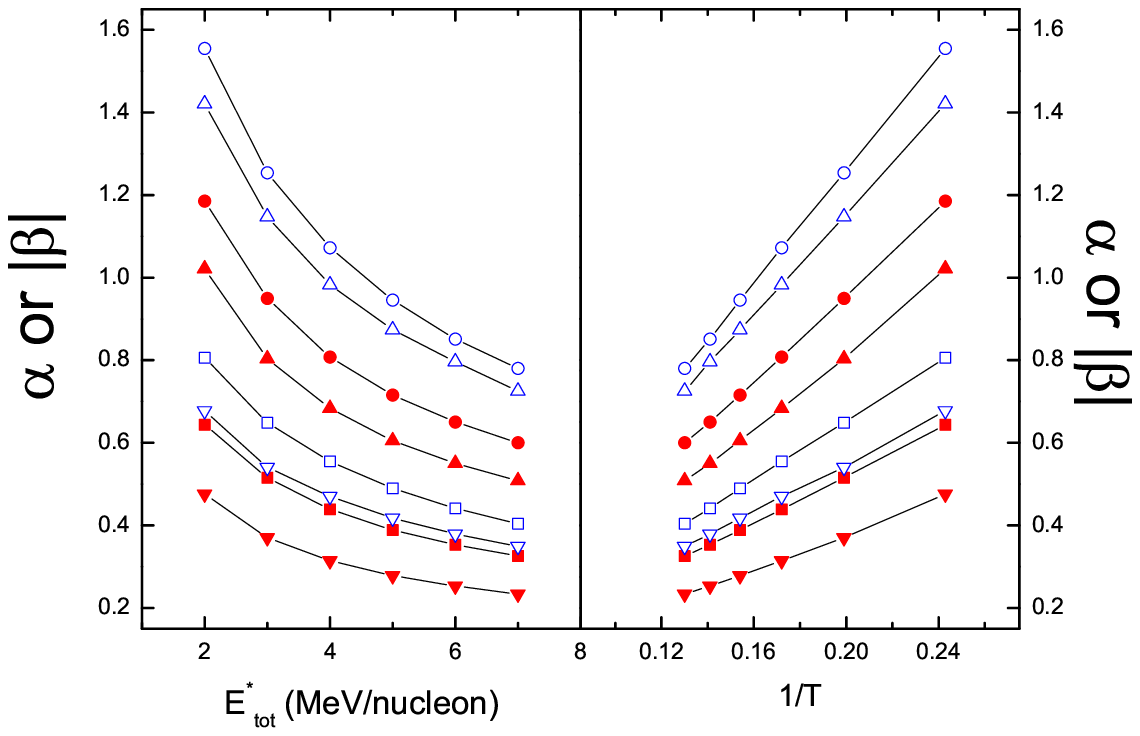}
\caption {\large (Color online) Same as Fig.2 but consider
 the surface entropy effect via inclusion of $a_S$
in the expression for the level density parameter $a$ [see Eq.
(6)]. }
\end{figure}

\newpage
\begin{figure}[htb]
\centering
\includegraphics[scale=1.0]{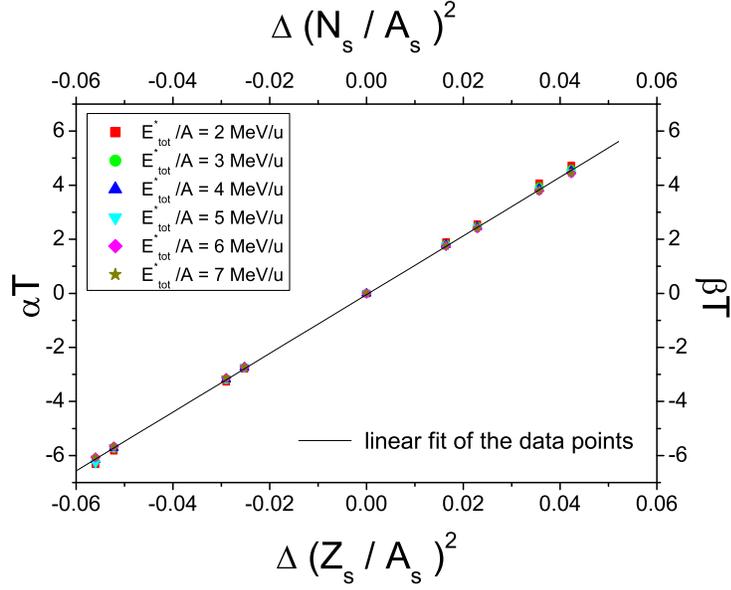}
\caption {\large (Color online) $\alpha \cdot T$ (positive parts)
and $\beta \cdot T$  (negative parts) as a function of
$(Z_s/A_s)^{2}_{1}$ - $(Z_s/A_s)^{2}_{2}$ or as a function of
$(N_s/A_s)^{2}_{1}$ - $(N_s/A_s)^{2}_{2}$ of the sources for four
source pairs, i.e., $Y_{A_{s}=175}$/$Y_{A_{s}=165}$,
$Y_{A_{s}=185}$/$Y_{A_{s}=165}$, $Y_{A_{s}=195}$/$Y_{A_{s}=175}$,
and $Y_{A_{s}=195}$/$Y_{A_{s}=185}$, at excitation energies of
$E^*_{tot}/A$ = 2, 3, 4, 5, 6, and 7 MeV/nucleon. Here the
calculations do not consider the surface entropy, i.e., not
including $a_S$ in the expression for the level density parameter
$a$ [see Eq. (6)]. }
\end{figure}

\newpage
\begin{figure}[htb]
\centering
\includegraphics[scale=1.0]{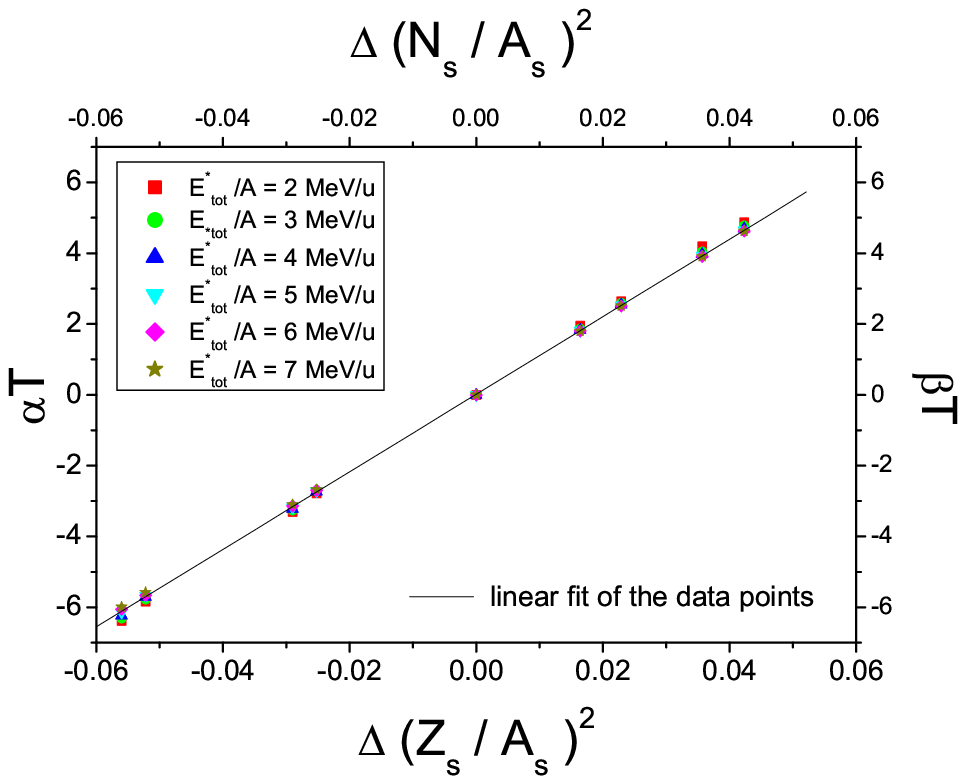}
\caption {\large (Color online) Same as Fig.5 but consider the
surface entropy effect via inclusion of $a_S$ in the expression
for the level density parameter $a$ [see Eq. (6)].}
\end{figure}

\end{document}